\documentclass[conference]{IEEEtran}
\IEEEoverridecommandlockouts

\usepackage{cite}
\usepackage{amsmath,amssymb,amsfonts}
\usepackage{algorithmic}
\usepackage{graphicx}
\usepackage{textcomp}
\usepackage{hyperref}
\usepackage{xcolor}
\def\BibTeX{{\rm B\kern-.05em{\sc i\kern-.025em b}\kern-.08em
    T\kern-.1667em\lower.7ex\hbox{E}\kern-.125emX}}
\begin{document}

\title{COVID-19 Detection System: A Comparative Analysis of System Performance Based on Acoustic Features of Cough Audio Signals}

\author{
  \IEEEauthorblockN{Asmaa Shati*\textsuperscript{1}\textsuperscript{2}, Ghulam Mubashar Hassan\textsuperscript{1}, Amitava Datta\textsuperscript{1}}
  \IEEEauthorblockA{
    \textsuperscript{1}Department of Computer Science and Software Engineering, University of Western Australia, Australia\\
    \textsuperscript{2}Department of Information Systems, King Khalid University, Saudi Arabia\\
    \ asmaa.shati@research.uwa.edu.au \\
    *Corresponding author 
  }
}



\maketitle

\begin{abstract}
 A wide range of respiratory diseases, such as cold and flu, asthma, and COVID-19, affect people's daily lives worldwide. In medical practice, respiratory sounds are widely used in medical services to diagnose various respiratory illnesses and lung disorders. The traditional diagnosis of such sounds requires specialized knowledge, which can be costly and reliant on human expertise. Recently, cough audio recordings have been used to automate the process of detecting respiratory conditions. This research aims to examine various acoustic features that enhance the performance of machine learning (ML) models in detecting COVID-19 from cough signals. This study investigates the efficacy of three feature extraction techniques, including Mel Frequency Cepstral Coefficients (MFCC), Chroma, and Spectral Contrast features, on two ML algorithms, Support Vector Machine (SVM) and Multilayer Perceptron (MLP), and thus proposes an efficient COVID-19 detection system. The proposed system produces a practical solution and demonstrates higher state-of-the-art classification performance with an AUC of 0.843 on the COUGHVID dataset and 0.953 on the Virufy dataset for COVID-19 detection.

\end{abstract}

\begin{IEEEkeywords}
machine learning, signal processing, feature extraction, COVID-19
\end{IEEEkeywords}

\section{Introduction}
Various respiratory diseases such as pneumonia, pertussis, COVID-19 and asthma, affect the respiratory system of human beings. In 2019, COVID-19 was identified as the source of the respiratory disease outbreak that began in China. 
COVID-19 affects the respiratory system, particularly the lungs and bronchial tubes, and its effects can gradually deteriorate and pose a threat to an individual’s life in certain circumstances. As of August, 2023, the pandemic had caused more than 768 million cases and 6.953 million confirmed deaths, making it one of the deadliest viruses in history \cite{covid192020coronavirus}. Therefore, researchers in different fields have attempted to determine the most effective strategy to combat the pandemic using new techniques, considering its severe medical and psychological consequences.

The universal approach for diagnosing COVID-19  requires human interaction and close contact. This approach seems insufficient to control the spread of the virus because of the grave threat that face-to-face test poses to medical personnel \cite{melek2022identifying}. Rapid Antigen Tests (RATs) can be used as an alternative approach. However, RATs require good-quality samples to be taken from the nose, particularly from a child's nose, which is challenging to obtain without parental involvement. The scientific and medical community has proposed using technology to detect COVID-19 from respiratory sounds, such as breath and cough recordings, as an alternative way of detecting the widespread infectious disease \cite{islam2022study}.

Machine Learning (ML) is a promising technique that has the potential to meaningfully employ medical data to enhance the quality of healthcare and minimize cost \cite{chaudhari2020virufy}. 
In the context of diagnosing respiratory diseases, developing machine-learning solutions would prove advantageous for preliminary screening and disease monitoring.

\begin{figure*}[h]
    \centering
        \includegraphics[width=0.4\linewidth]{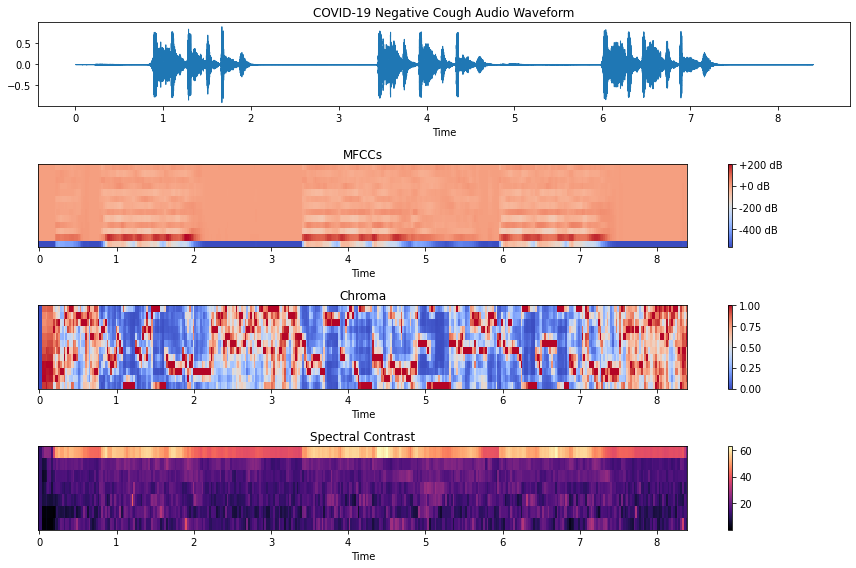}%
        \includegraphics[width=0.4\linewidth]{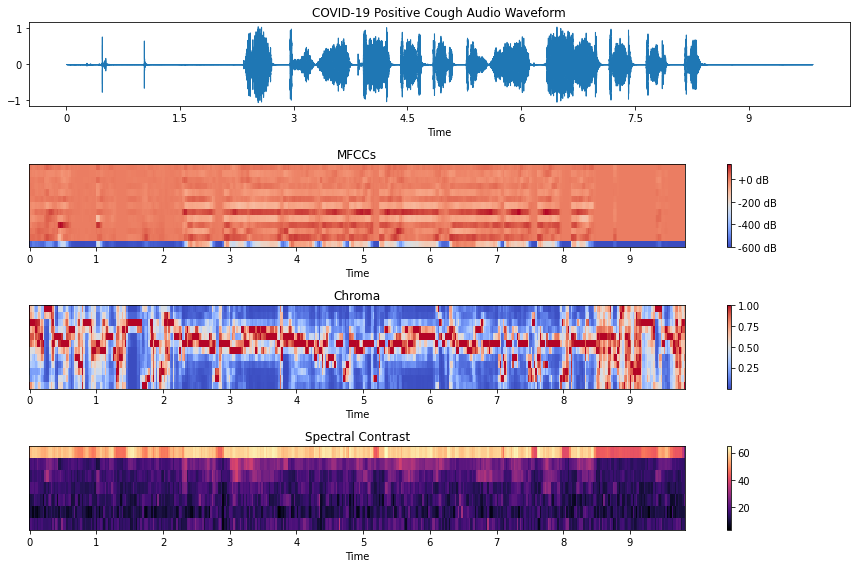}%
 \caption{(\textbf{Left}) COVID-19 positive cough sample (\textbf{Right}) COVID-19 negative cough sample}
    \label{cough-fig}
\end{figure*}

The primary contribution of this paper are summarized as follows:
\begin{itemize}
\item  A cost-effective integrated approach system of MFCC and an MLP is proposed to detect COVID-19 from cough signals.
\item  The proposed system is compared with state-of-the-art models, producing new benchmark results for COVID-19 detection.
\item The performance of the suggested system is assessed through six training and testing scenarios, which are evaluated using two publicly available datasets.
\item The efficiency of three sets of acoustic feature techniques in detecting COVID-19 is investigated using ablation studies.
\item  Unique to this work, the applicability of Spectral Contrast and Chroma features is demonstrated independently and in combination with other features in identifying COVID-19 from cough signals.
\end{itemize}
\section{Related WORK}
The usability of ML and respiratory sounds for detecting and screening COVID-19 has been explored in various studies \cite{despotovic2021detection,islam2022study, brown2020exploring,darici2022using, pahar2021covid,feng2021deep}. In \cite{feng2021deep}, the authors proposed a method for diagnosing COVID-19 from cough audio signals. They extracted time and frequency domain features from audio signals and then applied binary ML classifiers such as SVM, KNN, and RNN to classify them. They achieved an accuracy of 81.25\% and an AUC of 0.79 using RNN model.

Brown et al. \cite{brown2020exploring} collected cough and breathing data using a website and Android app to discriminate between healthy, COVID-19, and asthma patients. They extracted audio features from raw sound waveforms. Additionally, they employed pre-trained VGGish, a variant of the VGG model used for audio classification based on raw audio data input, to extract features from cough recordings. An AUC value of 0.80 was attained; however, the database used is not open source. Darici et al. \cite{darici2022using} applied two models to detect COVID-19 from cough on three different datasets, namely: COUGHVID, Coswara, and IATOS. The authors extracted several handcrafted features and used them as input to the SVM classifier, and a CNN model consisting of six convolutional layers. They reported an AUC of 0.807 for the SVM model and an AUC of 0.802 for the CNN model.

Pizzo and Esteban \cite{pizzo2021iatos} developed a CNN model consisting of two convolutional layers and one max-pooling layer for COVID-19 detection. The model was trained on the IATOS cough audio dataset, which was collected by the researchers using WhatsApp. They extracted the most relevant features from cough signals using Mel spectrogram. The overall accuracy of the proposed model is 86\%. The study suggested tailoring the model to different age groups and regions due to cultural and age variations in coughs. In \cite{sharma2020coswara}, the researchers collected a dataset of respiratory sounds called Coswara, including cough, breath, and voice sounds, and made them available to the public. The study extracted 28 acoustic features and trained the Random Forest algorithm with 30 trees, resulting in an accuracy of approximately 66.74\%.

In summary, the practical field of signal analysis for COVID-19 detection is developing continuously. Numerous research efforts have shown promising progress. However, there is a gap in the current literature involving COVID-19 detection from cough signals. Numerous studies do not explicitly specify the crucial features within cough signals that significantly contribute to improving model performance in detecting COVID-19. This matter serves as a motivation for us to conduct an investigation into the efficacy of various acoustic features that enhance the performance of the classification model in identifying COVID-19 from cough signals, which is effective in terms of training time, computational capability and accuracy. A pair of interpretable machine-learning models is trained using the extracted features from coughs. The comparative analysis of various models' explanations allows for gaining a better understanding of the relative importance of each set of extracted features.

\begin{figure*}[h]
\centering
\includegraphics[width=12 cm]{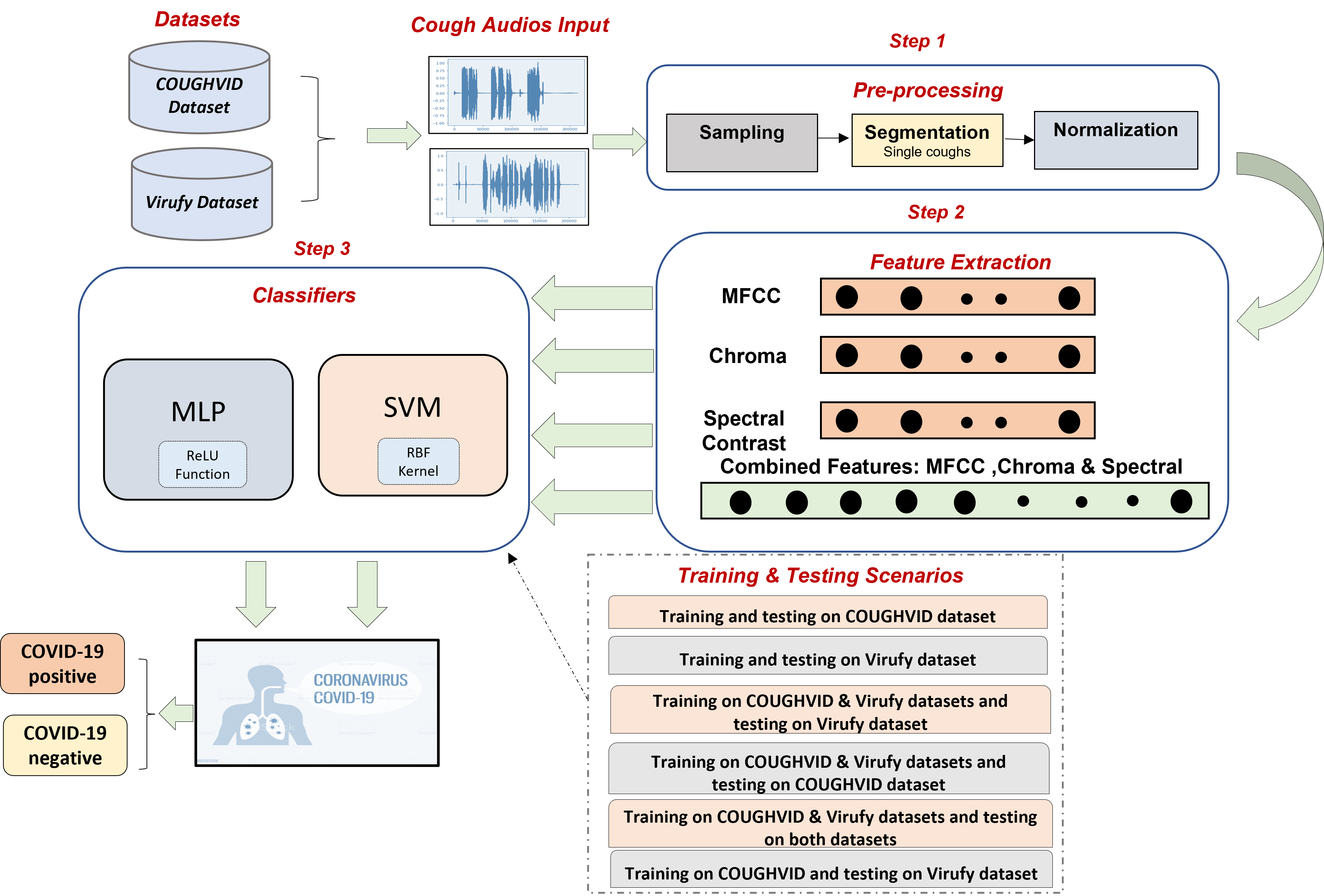}
\caption{Block diagram of the experimental procedure
\label{METHOD}}
\end{figure*}   
\section{Materials and Method}
Two cough sound samples were randomly selected to investigate the differentiation between coughs from individuals who tested positive for COVID-19 and those who tested negative, as illustrated in Fig \ref{cough-fig}. The figure illustrates a notable disparity in the signal of coughs between individuals who are COVID-19 positive and those who are negative. This implies that cough sounds have considerable potential as a data source for distinguishing between individuals who test positive or negative for COVID-19.

The experimental procedure used in this research is presented in Fig. \ref{METHOD}. It consists of three main steps: preprocessing, feature extraction, and classification. The first step, preprocessing, is important for preparing the data for subsequent analysis. This involves a sequential process of sampling, segmentation, and normalization. In the second step, features are extracted from the preprocessed audio data and fed into the classifiers. In the final step, MLP and SVM classifiers identify COVID-19 based on the features extracted from the cough signals.
\subsection{Datasets}
This study used two open-sourced datasets,  COUGHVID and Virufy. Both have cough audio signals and supplemental annotated files. 

COUGHVID dataset is the most extensive publicly accessible cough dataset for COVID-19 detection \cite{orlandic2021coughvid}. The audio recordings were collected from over 25,000 male and female participants. 
A part of this dataset over 2,800 cough recordings, was labeled by four expert physicians to diagnose medical issues. 

The Virufy dataset is clinically, reliable, and  accessible. Data were collected at a hospital under physician supervision from 16 individuals (10 males and 6 females) \cite{chaudhari2020virufy}.
The data were labeled based on  RT-PCR test results \cite{islam2022study}.
All recordings from both datasets were converted to wav format, and the sampling rate of audio files was set to 22 kHz using Librosa package.

We focused on high-quality recordings based on the annotated file information, such as the presence of coughing events and background noise. A total of 444 recordings were obtained from the COUGHVID and Virufy datasets, with 428 recordings from the former and 16 recordings from the latter, distributed across both classes (COVID-19 positive and COVID-19 negative). Every single recording had multiple coughs; these coughs were segmented as single coughs to ensure standardization for all recordings. The preprocessing yielded 1,953 segments, comprising 1,719 from the COUGHVID dataset and 234 from the Virufy dataset. Finally, the signals were normalized using the MinMaxScalar method to rescale data features between 0 and 1.

\subsection{Feature Extraction}
Three main acoustic features, MFCC, Chroma, and Spectral Contrast, were analyzed and applied in this study. The Short-Time Fourier Transform (STFT), which is a compromise between time and frequency-based representations, is required for extracting these features. The STFT identifies which frequencies are contained in the signal and at which point of time these frequencies occur \cite{shah2019chroma}.

\subsubsection{MFCC Features}
\begin{figure}[htbp]
\centering
\includegraphics[width=7 cm]{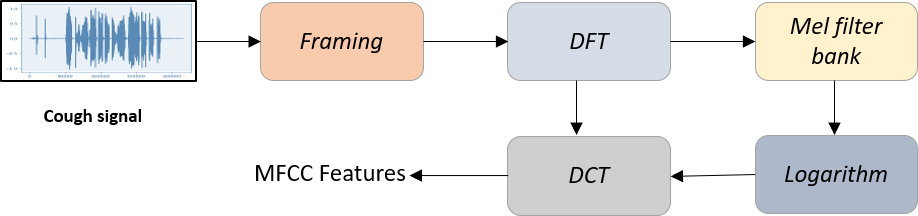}
\caption{MFCC feature extraction process
\label{MFCCFIG}}
\end{figure}

MFCC technique relies on human auditory perception of frequencies within the range of 20 to 20,000 Hz. 
The acronym MFCC comprises four distinct terms, namely Mel, frequency, cepstral, and coefficients, which summarize the fundamental principles of this particular approach. The MFCC procedure extracts the initial 13 coefficients from the signal, which have been determined to be satisfactory for representing speech or cough samples \cite{melek2022identifying, tzanetakis2002musical}. Our study extracted one set of the first 13 MFCC features from each cough recordings frame to detect COVID-19. The MFCC process is depicted in Fig. \ref{MFCCFIG}.

\subsubsection{Chroma Features}
Chroma differs from MFCC in capturing the type of information. While MFCC is primarily utilized to capture the spectral content from the signal, Chroma is employed to extract  the pitch content from the signal. It categorizes sounds into twelve distinct pitch classes \cite{shah2019chroma}. A 12-dimensional chroma feature represents the distribution of short-term energy in signals across twelve chroma bands \cite{MuellerKC05_ChromaFeatures_ISMIR, muller2011chroma}. In our experiments we extracted Chroma vector from each cough recording frame. The Chroma features $ V_{k}$ are mathematically represented in (1) \cite{islam2022study}. 
\begin{equation}
V_{k}=\sum_{n\in S_{k}} \frac{X_{i}(n)}{N_{k}}, k\in 0,....,11
\label{Equation 3}
\end{equation}
where the $V_{k}$
is a Chroma vector, $S_{k}$ is a subset of the frequencies, $N_{k}$ is the cardinality of $S_{k}$,
 $k$ is the pitch class and $i$ is the frame number.
 
\subsubsection{Spectral Contrast Features}  Spectral Contrast technique demonstrated successful application in audio genre classification as in \cite{west2005finding}. It considers the spectral peak, valley, and difference in each sub-band \cite{jiang2002music}. Each frame is composed of sub-bands. The energy contrast for each sub-band is calculated by comparing the mean energy at the top of the sub-band (peak energy) to that at the bottom of the sub-band (valley energy). We extracted a set of Spectral Contrast features from cough audio recordings to investigate its efficiency in detecting COVID-19.
The mathematical representation of Spectral Contrast $ SC_{k}$ is given in (2) \cite{jiang2002music}. 
\begin{equation}
SC_{k}={Peak_{k}} - {Valley_{k}}
\label{Equation 4}
\end{equation}
where $k$ indicates a sub-band, ${Peak_{k}}$ is the maximum value and ${Valley_{k}}$ is the minimum value in each sub-band.

\subsection{ Machine learning Models }
Two models, Multilayer Perceptron classifier (MLP) and  Support Vector Machine (SVM), were utilized in all the experiments. Both models were trained and tested using a random 80\%-20\% split. The choice of MLP classifier is motivated by its proven success in classifying musical genres, recognizing emotions in speech, and detecting COVID-19 from coughs \cite{kumar2021analysis,chowdhury2022machine, costa2004automatic}. The MLP classifier used in this study consists of four layers with sizes 300, 128, 64, and 2, followed by the rectified linear unit (RELU) activation function and ‘adam’ solver for weight optimization. The number of layers and hidden neurons were determined through multiple rounds of exploratory experiments where the  accuracy, F1-Score and the Area Under the Curve (AUC) were used as criteria for assessment.

In addition, the selection of SVM classifier is suggested because of its promising results in audio classification tasks \cite{melek2022identifying, Kerkeni19, hemdan2022cr19, erdougan2021covid}. In our experiments, we utilized the radial basis function (RBF) kernel, which performed better than  other kernel functions, such as linear, polynomial, and sigmoid functions.
We also optimized the regularization hyper-parameters, gamma ($\gamma$) and (C), as they are known to have a significant impact on the performance of SVM model \cite{geron2022hands}. A range of gamma values from 0 to 1 was tested by step 0.2 to optimize the gamma parameter in the SVM classifier. The highest accuracy was attained at gamma = 0.4. The value of C was optimized to 20 after experimenting with a range of values between 1 and 100 with step 5. Table \ref{hyper} shows the details of hyper-parameters optimization for both MLP and SVM classifiers.

\begin{table*}[ht]
\centering
\caption{Hyper-parameters optimization for MLP and SVM classifiers}\label{hyper}
\begin{tabular}{cll}
\hline
\textbf{Classifier}           & \multicolumn{1}{c}{\textbf{Hyper-parameters}} & \multicolumn{1}{c}{\textbf{Range}}                                                                                                                                     \\  \hline
{\textbf{MLP}} & Number of hidden layers                       & 1 to 4 \\
                              & Hidden layers sizes                           & \begin{tabular}[c]{@{}l@{}}(300), (128), (64), (300,300), (128,128) (64,64), (300,128), \\ (300,64), (128,64), (300, 128, 64), (300, 128, 64, 2)\end{tabular} \\
                              & Solver                                        & lbfgs, sgd, adam                                                                                                                                                       \\
                              & Learning rate                                 & constant, invscaling, adaptive                                                                                                                                         \\
                              & Activation function                           & identity, logistic, tanh, relu                                                                                                                                         \\ \hline
{\textbf{SVM}} & Kernel                                        & linear, poly, rbf, sigmoid                                                                                                                                             \\
                              & Gamma                                         & 0 to 1, step = 0.02                                                                                                                                                    \\
                              & C                                             & 1 to 100, step =5                                                                                                                                                      \\ \hline
\end{tabular}
\end{table*}
\section{Results}

In this study, we tested our proposed method for detecting COVID-19 on two datasets: COUGHVID and Virufy, by using different evaluation metrics to assess the performance of the classification models. These metrics include accuracy, precision, recall,  F1-score, and the area under the curve (AUC).

Individual acoustic features proposed in this work were analyzed to determine how well they contributed to a model's overall COVID-19 detection performance. In all experiments, both SVM and MLP classifiers were implemented for each extracted feature set: MFCC, Chroma, Spectral Contrast, and then for a combined feature set. This procedure was conducted to assess the applicability of the extracted features in detecting COVID-19 from cough audio recordings. We proposed six training and testing scenarios to train and test the classifiers, as depicted in Fig. \ref{METHOD}.

We applied four techniques to extract acoustic features, as shown in Fig. \ref{METHOD}, and evaluated their efficacy in detecting COVID-19 from coughs. The first technique involved extracting thirteen MFCC features from cough audio signals and feeding them into the SVM and MLP classifiers. 
Next, a set of twelve Chroma features were extracted and passed to the classifiers in a similar manner. The third technique involved extracting a set of Spectral Contrast features, which were employed as input for the classification models. In the final technique, we combined all the extracted features and utilized them as input for our classification models.

Tables \ref{R1} and \ref{R2} show the detailed performance metrics for training and testing on COUGHVID and Virufy datasets, respectively, applying the extraction techniques of the proposed features on both datasets. Looking at these tables, it is clear that the best performance comes from the MLP classifier when fed only with MFCC features in both datasets. The MLP classifier achieved 84.30\% accuracy for COUGHVID dataset and 96.10\% accuracy for Virufy dataset, whereas the SVM classifier achieved 83.21\% accuracy for COUGHVID and 91.02\% accuracy for Virufy dataset. The combination of the proposed features in this study yields a satisfactory performance, which indicates that utilizing all of these features together is useful in analyzing cough signals. Specifically, this is the case for the MLP classifier, which attained 83.07\% accuracy for COUGHVID dataset and 94.13\%  accuracy for Virufy dataset. 

In addition, it can be observed from Tables \ref{R1} and \ref{R2} that there is little difference between the performance of the SVM and MLP classifiers when Chroma features are utilized as input. The variation between the two models is one percent. The model's performance on Virufy dataset is noteworthy, achieving an accuracy of approximately 81.22\% for the MLP classifier. However, its accuracy on COUGHVID dataset is comparatively lower, with the MLP classifier achieving an accuracy of approximately 69.14\%. The potential cause for this difference could be attributed to the presence of noise that impacted the crowdsourced cough recordings presented in the COUGHVID dataset, in contrast to the cleaned recordings found in Virufy dataset. The spectral characteristics of sound are heavily influenced by the frequency response of the microphone used to capture it \cite{muller2007information}. In the case of recordings obtained through crowdsourcing, such as those in COUGHVID dataset, the microphone's ability to capture very low frequencies is frequently inadequate.

Furthermore, it can be seen from Tables \ref{R1} and \ref{R2} that the Spectral Contrast features yield varying results. Specifically, the MLP classifier attains accuracy of 70.03 \% and 85.36\%  for COUGHVID and Virufy, respectively. The variation in the model's performance on the two datasets could be due to the disparity in the quality of cough recordings in Virufy and COUGHVID datasets, similar to the potential impact of Chroma features.

\begin{table*}[h]
\centering
\caption{The evaluation metrics of training and testing MLP and SVM classifiers on \textbf{COUGHVID} dataset using MFCC, Chroma, Spectral Contrast features and combination of them. The highest values of AUC and Accuracy are written in bold. \label{R1}}

\begin{tabular}{ccccccc}
\hline
\textbf{Model}                &\textbf{Class}    &\textbf{Precision} & \textbf{Recall} & \textbf{F1-Score} &\textbf{AUC}                    & \textbf{Acc}                   \\ \hline
\multicolumn{7}{c}{\textbf{MFCC Features}}                                                                                \\ \hline
{\textbf{MLP}} & {COVID-19 positive} & {80.0}      & {87.1}   & {83.4}     & {{\textbf{0.843}}} & {{\textbf{84.30}}} \\
                     & {COVID-19 negative}  & {{88.2}}     & {82.0}  & {85.0}    &                        &                        \\
{\textbf{SVM}} & {COVID-19 positive} & {87.1}      & {82.3}  & {84.6}     & {{0.834}} & {{83.21}} \\
                     & {COVID-19 negative}  & {80.3}      & {86.4}  & {83.2}     &                        &                        \\ \hline
\multicolumn{7}{c}{\textbf{Chroma Features}}                                                                               \\ \hline
{\textbf{MLP}} & {COVID-19 positive} & {71.1}      &{68.2}   & {70.0}     & {{0.691}} & {{69.14}} \\
                     & {COVID-19 negative}  & {67.3}     & {70.4}   & {69.0}     &                        &                        \\
{\textbf{SVM}} & {COVID-19 positive} & {73.2}      & {67.3}   & {70.1}     & {{0.685}} & {{68.51}} \\
                     &{COVID-19 negative}  & {64.0}      & {70.5}   & {67.1}     &                        &                        \\ \hline
\multicolumn{7}{c}{\textbf{Spectral Contrast Features}}                                                                   \\ \hline
{\textbf{MLP}} & {COVID-19 positive} & {66.2}      & {71.0}   & {{69.5}}     & {{0.697}} & {{70.03}} \\
                     & {COVID-19 negative}  & {74.1}      & {68.3}   & {71.1}     &                        &                        \\
{\textbf{SVM}} & {COVID-19 positive} & {69.4}      & {65.0}   & {67.1}     & {{0.661}} & {{66.17}} \\
                     & {COVID-19 negative}  & {64.0}      & {67.1}   & {65.5}     &                        &                        \\ \hline
\multicolumn{7}{c}{\textbf{Combined Features (MFCC, Chroma and Spectral Contrast)}}                                        \\ \hline
{\textbf{MLP}} & {COVID-19 positive} & {81.0}      & {85.2}   & {83.0}     & {{0.829}} & {{83.07}} \\
                     & {COVID-19 negative}  & {85.3}      & {81.0}   & {83.1}     &                        &                        \\
{\textbf{SVM}} & {COVID-19 positive} & {87.2}      & {75.4}   & {81.0}     & {{0.776}} & {\footnotesize{78.22}} \\
                     & {COVID-19 negative}  & {66.1}      & {86.3}   & {74.8}     &                        &                        \\ \hline
\end{tabular}
\end{table*}
\begin{table*}[h]
\centering
\caption{The evaluation metrics of training and testing MLP and SVM classifiers on \textbf{Virufy} dataset using MFCC, Chroma, Spectral Contrast features and combination of them. The highest values of AUC and Accuracy are written in bold.\label{R2}}
\begin{tabular}{ccccccc}
\hline
\textbf{Model}                & \textbf{Class}    & \textbf{Precision} & \textbf{Recall} & \textbf{F1-Score} &\textbf{AUC}                    & \textbf{Acc}                   \\  \hline
\multicolumn{7}{c}{\textbf{MFCC Features}}                                                                                \\ \hline
{\textbf{MLP}} & 
{COVID-19 positive} & {94.0}      & {94.1}   & {94.0}     & {{\textbf{0.953}}} & {{\textbf{96.10}}} \\
                     & {COVID-19 negative}  & {97.1}      & {97.2}  & {97.2}     &                        &                        \\
{\textbf{SVM}} & {COVID-19 positive} & {88.2}      & {88.3}  & {88.2}     & {{0.907}} & {{91.02}} \\
                     & {COVID-19 negative}  & {93.3}      & {93.0}   & {93.1}     &                        &                        \\ \hline
\multicolumn{7}{c}{\textbf{Chroma Features}}                                                                               \\ \hline
{\textbf{MLP}} & {COVID-19 positive} & {76.1}      &{72.0}   & {74.0}     & {{0.799}} & {{81.22}} \\
                     & {COVID-19 negative}  & {83.0}     & {86.4}   & {84.7}     &                        &                        \\
                     {\textbf{SVM}} & {COVID-19 positive} & {65.2}      & {85.3}   & {74.0}     & {{0.790}} & {{82.26}} \\
                     & {COVID-19 negative}  & {93.4}      & {82.1}   & {87.4}     &                        &                        \\ \hline
\multicolumn{7}{c}{\textbf{Spectral Contrast Features}}                                                                   \\ \hline
{\textbf{MLP}} & {COVID-19 positive} & {88.1}      & {75.0}   & {{81.0}}     & {{0.857}} & {{85.36}} \\
                     & {COVID-19 negative}  & {83.0}      & {93.2}   & {87.8}     &                        &                        \\
{\textbf{SVM}} & {COVID-19 positive} & {65.3}      & {73.4}   & {69.1}     & {{0.756}} & {{78.71}} \\
                     & {COVID-19 negative}  & {87.2}      & {81.1}   & {84.0}     &                        &                        \\ \hline
                     \multicolumn{7}{c}{\textbf{Combined Features (MFCC, Chroma and Spectral Contrast)}}                                        \\ \hline
{\textbf{MLP}} & {COVID-19 positive} & {88.2}      & {94.3}   & {91.1}     & {{0.924}} & {{94.13}} \\
                     & {COVID-19 negative}  & {97.0}      & {94.0}   & {95.5}     &                        &                        \\
{\textbf{SVM}} & {COVID-19 positive} & {82.1}      & {93.5}   & {87.4}     & {{0.895}} & {{91.40}} \\
                     & {COVID-19 negative}  & {97.1}      & {91.1}   & {94.0}     &                        &                        \\ \hline
\end{tabular}
\end{table*}
\subsection{Ablation Studies}
To assess the effectiveness of the proposed technique, the experiments were repeated using the remaining four training and testing scenarios, as illustrated in Fig. \ref{METHOD}.  Table \ref{Ablation} lists the top performance metrics obtained from ablation experiments. The findings of our experiments shown in Table \ref{Ablation} demonstrate that MFCC features provide the highest performance measures compared to other features in terms of area under the curve (AUC), accuracy, and F1-score across all experimental settings. We observed that training our data on samples from both datasets and testing them separately yielded a varying average F1-score of about 93.45\% when testing on Virufy dataset and 78.6\% when testing on COUGHVID dataset. However, training and testing on samples from both datasets resulted in a reasonable average F1-score of about 83\%. This would suggest that the model possesses the ability to differentiate between coughs, owing to its acquisition of a greater variety of data characteristics from both datasets. As expected, the model exhibited a slightly dropped average F1-score of approximately 71.0\% when trained on COUGHVID dataset and tested on Virufy dataset, partially due to the differences in audio recordings characteristics between the two datasets.
\begin{table*}[h]
\centering
\caption{The best evaluation metrics achieved in the remaining four scenarios, specifically when using \textbf{MFCC} features with MLP classifier, are summarized below.\label{Ablation}}
\begin{tabular}{cccc}
\hline

\textbf{Experimental Scenario}                                                                                     & \textbf{AUC}   & \textbf{Accuracy} & \textbf{Avg F1-score} \\ \hline

\begin{tabular}[c]{@{}c@{}} Training on COUGHVID \& Virufy datasets \\  and testing on Virufy dataset\end{tabular}   &\footnotesize 0.899 &\footnotesize 91.0    \footnotesize & \footnotesize 93.45        \\ \hline
\begin{tabular}[c]{@{}c@{}} Training on COUGHVID \&  Virufy datasets \\  and testing on COUGHVID dataset\end{tabular} &\footnotesize 0.789 & \footnotesize79.0     & \footnotesize78.6         \\ \hline
\begin{tabular}[c]{@{}c@{}} Training on COUGHVID \& Virufy datasets \\  and testing on both\end{tabular}             &\footnotesize 0.833 &\footnotesize 83.24    &\footnotesize 83.0         \\ \hline
\begin{tabular}[c]{@{}c@{}} Training on COUGHVID dataset and testing \\  on Virufy dataset\end{tabular}              & \footnotesize 0.639 &\footnotesize 64.11    & \footnotesize71.0         \\ \hline

\end{tabular}
\end{table*}
\begin{table}
    \centering
    \caption{The training and runtime performance of the proposed method on both COUGHVID and Virufy datasets is presented in seconds}
    \label{Runtime}
    \begin{tabular}{ccc}
    \hline
         \textbf{Dataset} & \textbf{Training Time} & \textbf{Runtime}\\ \hline
        COUGHVID & 2.650  &  0.003   \\  \hline
        Virufy& 0.574  & 0.001 \\ \hline
    \end{tabular}
\end{table}
Taking into account the overall outcomes, the classification results of cough signals of COVID-19 seem more efficient when using MLP classifier with MFCC technique as MFCC yields better performance compared to Chroma and Spectral Contrast features. Using a combination of acoustic features gives promising results in most cases. However, in all the settings, MFCC technique achieves superior results for cough audio classification. Table \ref{Runtime} shows the training and runtime performance of the proposed method on both COUGHVID and Virufy datasets, indicating that it would be valuable for COVID-19 classification tasks from cough signals without requiring excessive complexity or computational time.

\section{Discussion}
 
In this study, we analyzed cough audios using several acoustic feature extraction methods and compared their performance. Our analysis facilitates the performance comparison of the suggested features, including MFCC, Chroma, and Spectral Contrast, when studying cough signals.
To the best of our knowledge, this is the first study in this field to examine different proposed acoustic feature utilizing two machine learning classifiers across various training and testing scenarios as depicted in Tables  \ref{R1}, \ref{R2} and \ref{Ablation}. This procedure was conducted to shed light on these features and assess their suitability for analyzing cough signals, paving the way for further investigation and aiding in selecting appropriate acoustic features, which is a crucial step in distinguishing between various cough types and respiratory diseases.

The present study exploits two machine learning models to assess their performance when fed with a limited and an expanded set of acoustic features. The MLP classifier was found to be more beneficial than the SVM classifier in the majority of experiments when analyzing cough sound signals. The proposed MLP classifier trained with MFCC features outperforms the other classifier trained with other features, such as Chroma and Spectral Contrast, in detecting COVID-19. This can be explained by the fact that MLP can learn intricate patterns in data, which would be valuable when working with cough signals. Such signals can exhibit complex patterns that may be challenging for simpler classifiers like SVM. Moreover, MFCC features are exceptionally efficient in speech recognition tasks, as they capture significant characteristics from audio signals, which is similarly applicable to cough signals. These characteristics could include differences in cough pitch, duration, and spectral content for cough classification.
Melek \cite{melek2022identifying}  has shown MFCC technique's success when comparing two feature extraction methods, MFCC and STFT. This provides further evidence in support of what we have observed in our work, which is that MFCC-based features are highly recommended for extracting features from cough audio signals to detect respiratory diseases.

It is noteworthy that selecting suitable acoustic features for classifying cough audio signals is a developing research area.
We observed that Spectral Contrast and Chroma techniques had yet to be exclusively explored in the domain of cough sounds, as they were combined with other features. Therefore, we examine their efficacy independently and with other acoustic features, providing practical insights into the use of such features to analyze cough signals. The study confirms the applicability of Spectral Contrast and Chroma features in detecting COVID-19 from coughs.

\begin{table*}[htbp]
\centering
\caption{\label{work}Comparison of the proposed COVID-19 detection model against existing state-of-the-art work}

\begin{tabular}{llllll}
\hline \\
\textbf{Study}              & \textbf{Model}                                             & \textbf{Features}                                                                                                       & \textbf{Acc}                                                                   & \textbf{AUC}                                                                   & \textbf{Dataset}                                                          \\ \\ \hline
Ponomarchuk et al. \cite{ponomarchuk2022project} & \begin{tabular}[c]{@{}l@{}}CNN\\ GB\end{tabular}  & STFT and Mel-spectogram                                                                                        & -                                                                     & \begin{tabular}[c]{@{}l@{}}0.630\\ 0.592\end{tabular}                 & COUGHVID                                                         \\
Chaudhari et al. \cite{chaudhari2020virufy}   & DNN                                               & MFCC and Spectogram                                                                                            & -                                                                     & 0.771                                                                 & COUGHVID/Coswara                                                 \\
Darici et al. \cite{darici2022using}  & \begin{tabular}[c]{@{}l@{}}CNN\\ SVM\end{tabular} & \begin{tabular}[c]{@{}l@{}}MFCC, Delta, Spectral Centroid\\ Spectral Roll-off and RMS\end{tabular}             & \begin{tabular}[c]{@{}l@{}}75.0\\ 79.0\end{tabular}                   & \begin{tabular}[c]{@{}l@{}}0.802\\ 0.807\end{tabular}                 & \begin{tabular}[c]{@{}l@{}}Virufy/COUGHVID/ IATOS\end{tabular} \\
Feng et al. \cite{feng2021deep}        & RNN                                               & \begin{tabular}[c]{@{}l@{}}ZCR, Energy, Entropy of Energy,\\ Spectral measurements and MFCC\end{tabular}        & 81.25                                                                 & 0.79                                                                  & Virufy/Coswara                                                   \\
Islam et al. \cite{islam2022study}      & DNN                                               & \begin{tabular}[c]{@{}l@{}}Time domain\\ Frequency domain\\ Mixed domain\end{tabular}                          & \begin{tabular}[c]{@{}l@{}}89.2\\ {\textbf{97.2}}\\ 93.8\end{tabular}            & \begin{tabular}[c]{@{}l@{}}-\\ -\\ -\end{tabular}                     & Virufy                                                           \\
Melek \cite{melek2022identifying}          & SVM                                               & MFCC                                                                                                           & 95.86                                                                 & -                                                                     & Virufy                                                           \\ \hline
Proposed Model     & {\textbf{MLP}}                                               & \begin{tabular}[c]{@{}l@{}}MFCC\\ Chroma\\ Spectral Contrast\\ MFCC, Chroma and Spectral contrast\end{tabular} & \begin{tabular}[c]{@{}l@{}}96.10\\ 81.22\\ 85.36\\ 94.13\end{tabular} & \begin{tabular}[c]{@{}l@{}}{\textbf{0.953}}\\ 0.80\\ 0.857\\ 0.924\end{tabular}  & {\textbf{Virufy}}                                                           \\ \cline{3-6} 
                   &                                                   & \begin{tabular}[c]{@{}l@{}}MFCC\\ Chroma\\ Spectral Contrast\\ MFCC, Chroma and Spectral contrast\end{tabular} & \begin{tabular}[c]{@{}l@{}}{\textbf{84.30}}\\ 69.14\\ 70.03\\ 83.07\end{tabular} & \begin{tabular}[c]{@{}l@{}}{\textbf{0.843}}\\ 0.691\\ 0.697\\ 0.829\end{tabular} & {\textbf{COUGHVID}} \\ \hline                                                       
\end{tabular}
\end{table*}
To demonstrate the effectiveness of our technique, we compared our results with those of previous studies. Since our study aims to contribute to the ongoing efforts in detecting COVID-19 through coughs, we tested the proposed model using the COUGHVID and Virufy datasets. The results are presented in Table \ref{work}.
This table shows the results of our proposed scenarios, one and two, where the training and testing processes were carried out independently on the COUGHVID and Virufy datasets. The purpose was to compare these results with findings on the same datasets from other studies.

It can be observed from Table \ref{work} that the suggested methods by \cite{ponomarchuk2022project}, applied to the COUGHVID dataset, exhibit relatively poor performance when compared to the results achieved by our proposed MLP model. The study \cite{ponomarchuk2022project} employed STFT and Melspectogram features as inputs for CNN and gradient-boosted models. However, it is worth noting that this feature selection strategy may be inadequate for effectively discerning cough patterns. Furthermore, we notice that little work has been done on the COUGHVID dataset. We found two studies \cite{chaudhari2020virufy,darici2022using} that used some of its samples along with other datasets, which makes it challenging to compare their findings with ours. Nevertheless, even in these two studies, where the utilization of MFCC and other acoustic features in conjunction with deep learning models was evident, it is apparent that their approaches have not yet yielded performance improvements in detecting COVID-19 coughs to the extent achieved by our proposed model.

Furthermore, we compare our work using Virufy dataset with other studies which used the same dataset in their work. Our MLP model provides better accuracy compared to the work done by \cite{melek2022identifying, feng2021deep}. Study \cite{islam2022study} shows different accuracy when extracting time-domain, frequency domain, and mixed features, and it gives the highest accuracy with frequency domain features. However, our suggested approach yielded a high AUC of 0.953, which was not provided in the same study.

Finally, it is essential to acknowledge certain constraints that may exist. We observed differences in the performance of our model when testing it on two distinct datasets collected in an uncontrolled environment (COUGHVID dataset) and in a controlled environment (Virufy dataset). We attribute the presence of background noise and variations in the sources from which the audio recordings of COUGHVID dataset were collected to this matter. To address this issue, we included high-quality recordings verified in the annotated file and a high Signal-to-Noise Ratio (SNR) level provided by the authors in \cite{orlandic2021coughvid}. In the future, the focus can be directed towards reducing background noise by implementing noise filtering techniques.

\section{Conclusions}
This study proposes a cost-effective pre-screening system for detecting COVID-19 from cough signals. The system is developed based on a comparative analysis of system performance utilizing three acoustic feature sets: MFCC, Chroma, and Spectral Contrast extracted from cough audio signals in a pair of machine learning models. We conduct comprehensive simulations with six training and testing scenarios across two open-sourced datasets to train and evaluate the classification models. The findings of this study demonstrate that our integrated approach system, combining MFCC features with MLP models, provides an efficient solution for detecting COVID-19 from cough recordings without the need for excessive computational time or resources. Additionally, we explore the potential use of Spectral Contrast and Chroma, two other acoustic feature sets, in identifying COVID-19 from cough samples, thus opening up possibilities for enhancing the utilization of these features in the field of cough analysis.


\bibliographystyle{ieeetr}

\begin{thebibliography}{10}

\bibitem{covid192020coronavirus}
W.~H. Organization, ``Who coronavirus disease (covid-19) dashboard.'' \url{https://covid19.who.int/}, 2020.
\newblock Accessed on August 23, 2023.

\bibitem{melek2022identifying}
N.~Melek~Manshouri, ``Identifying covid-19 by using spectral analysis of cough recordings: a distinctive classification study,'' {\em Cognitive Neurodynamics}, vol.~16, no.~1, pp.~239--253, 2022.

\bibitem{islam2022study}
R.~Islam, E.~Abdel-Raheem, and M.~Tarique, ``A study of using cough sounds and deep neural networks for the early detection of covid-19,'' {\em Biomedical Engineering Advances}, vol.~3, p.~100025, 2022.

\bibitem{chaudhari2020virufy}
G.~Chaudhari, X.~Jiang, A.~Fakhry, A.~Han, J.~Xiao, S.~Shen, and A.~Khanzada, ``Virufy: Global applicability of crowdsourced and clinical datasets for ai detection of covid-19 from cough,'' {\em arXiv preprint arXiv:2011.13320}, 2020.

\bibitem{despotovic2021detection}
V.~Despotovic, M.~Ismael, M.~Cornil, R.~Mc~Call, and G.~Fagherazzi, ``Detection of covid-19 from voice, cough and breathing patterns: Dataset and preliminary results,'' {\em Computers in Biology and Medicine}, vol.~138, p.~104944, 2021.

\bibitem{brown2020exploring}
C.~Brown, J.~Chauhan, A.~Grammenos, J.~Han, A.~Hasthanasombat, D.~Spathis, T.~Xia, P.~Cicuta, and C.~Mascolo, ``Exploring automatic diagnosis of covid-19 from crowdsourced respiratory sound data,'' {\em arXiv preprint arXiv:2006.05919}, 2020.

\bibitem{darici2022using}
E.~Darici, N.~Rasmussen, J.~Xiao, G.~Chaudhari, A.~Rajput, P.~Govindan, M.~Yamaura, L.~Gomezjurado, A.~Khanzada, M.~Pilanci, {\em et~al.}, ``Using deep learning with large aggregated datasets for covid-19 classification from cough,'' {\em arXiv preprint arXiv:2201.01669}, 2022.

\bibitem{pahar2021covid}
M.~Pahar, M.~Klopper, R.~Warren, and T.~Niesler, ``Covid-19 cough classification using machine learning and global smartphone recordings,'' {\em Computers in Biology and Medicine}, vol.~135, p.~104572, 2021.

\bibitem{feng2021deep}
K.~Feng, F.~He, J.~Steinmann, and I.~Demirkiran, ``Deep-learning based approach to identify covid-19,'' in {\em SoutheastCon 2021}, pp.~1--4, IEEE, 2021.

\bibitem{pizzo2021iatos}
D.~T. Pizzo, S.~Esteban, and M.~Scetta, ``Iatos: Ai-powered pre-screening tool for covid-19 from cough audio samples,'' {\em arXiv preprint arXiv:2104.13247}, 2021.

\bibitem{sharma2020coswara}
N.~Sharma, P.~Krishnan, R.~Kumar, S.~Ramoji, S.~R. Chetupalli, P.~K. Ghosh, S.~Ganapathy, {\em et~al.}, ``Coswara--a database of breathing, cough, and voice sounds for covid-19 diagnosis,'' {\em arXiv preprint arXiv:2005.10548}, 2020.

\bibitem{orlandic2021coughvid}
L.~Orlandic, T.~Teijeiro, and D.~Atienza, ``The coughvid crowdsourcing dataset, a corpus for the study of large-scale cough analysis algorithms,'' {\em Scientific Data}, vol.~8, no.~1, pp.~1--10, 2021.

\bibitem{shah2019chroma}
A.~Shah, M.~Kattel, A.~Nepal, and D.~Shrestha, ``Chroma feature extraction,'' {\em Chroma Feature Extraction using Fourier Transform}, 2019.

\bibitem{tzanetakis2002musical}
G.~Tzanetakis and P.~Cook, ``Musical genre classification of audio signals,'' {\em IEEE Transactions on speech and audio processing}, vol.~10, no.~5, pp.~293--302, 2002.

\bibitem{MuellerKC05_ChromaFeatures_ISMIR}
M.~M{\"u}ller, F.~Kurth, and M.~Clausen, ``Audio matching via chroma-based statistical features,'' in {\em Proceedings of the 6th International Conference on Music Information Retrieval ({ISMIR})}, pp.~288--295, 2005.

\bibitem{muller2011chroma}
M.~M{\"u}ller and S.~Ewert, ``Chroma toolbox: Pitch, chroma, cens, crp,'' in {\em Proc. ISMIR}, 2011.

\bibitem{west2005finding}
K.~West and S.~Cox, ``Finding an optimal segmentation for audio genre classification.,'' in {\em ISMIR}, pp.~680--685, 2005.

\bibitem{jiang2002music}
D.-N. Jiang, L.~Lu, H.-J. Zhang, J.-H. Tao, and L.-H. Cai, ``Music type classification by spectral contrast feature,'' in {\em Proceedings. IEEE International Conference on Multimedia and Expo}, vol.~1, pp.~113--116, IEEE, 2002.

\bibitem{kumar2021analysis}
S.~Kumar and S.~Thiruvenkadam, ``An analysis of the impact of spectral contrast feature in speech emotion recognition.,'' {\em Int. J. Recent Contributions Eng. Sci. IT}, vol.~9, no.~2, pp.~87--95, 2021.

\bibitem{chowdhury2022machine}
N.~K. Chowdhury, M.~A. Kabir, M.~M. Rahman, and S.~M.~S. Islam, ``Machine learning for detecting covid-19 from cough sounds: An ensemble-based mcdm method,'' {\em Computers in Biology and Medicine}, vol.~145, p.~105405, 2022.

\bibitem{costa2004automatic}
C.~H.~L. Costa, J.~D. Valle, and A.~L. Koerich, ``Automatic classification of audio data,'' in {\em 2004 IEEE International Conference on Systems, Man and Cybernetics (IEEE Cat. No. 04CH37583)}, vol.~1, pp.~562--567, IEEE, 2004.

\bibitem{Kerkeni19}
L.~Kerkeni, Y.~Serrestou, M.~Mbarki, K.~Raoof, M.~A. Mahjoub, and C.~Cleder, ``Automatic speech emotion recognition using machine learning,'' in {\em Social Media and Machine Learning} (A.~Cano, ed.), ch.~2, Rijeka: IntechOpen, 2019.

\bibitem{hemdan2022cr19}
E.~E.-D. Hemdan, W.~El-Shafai, and A.~Sayed, ``Cr19: A framework for preliminary detection of covid-19 in cough audio signals using machine learning algorithms for automated medical diagnosis applications,'' {\em Journal of Ambient Intelligence and Humanized Computing}, pp.~1--13, 2022.

\bibitem{erdougan2021covid}
Y.~E. Erdo{\u{g}}an and A.~Narin, ``Covid-19 detection with traditional and deep features on cough acoustic signals,'' {\em Computers in Biology and Medicine}, vol.~136, p.~104765, 2021.

\bibitem{geron2022hands}
A.~G{\'e}ron, {\em Hands-on machine learning with Scikit-Learn, Keras, and TensorFlow}.
\newblock " O'Reilly Media, Inc.", 2022.

\bibitem{muller2007information}
M.~M{\"u}ller, {\em Information retrieval for music and motion}, vol.~2.
\newblock Springer, 2007.

\bibitem{ponomarchuk2022project}
A.~Ponomarchuk, I.~Burenko, E.~Malkin, I.~Nazarov, V.~Kokh, M.~Avetisian, and L.~Zhukov, ``Project achoo: a practical model and application for covid-19 detection from recordings of breath, voice, and cough,'' {\em IEEE Journal of Selected Topics in Signal Processing}, vol.~16, no.~2, pp.~175--187, 2022.

\end{thebibliography}

\end{document}